\documentclass[
 aps, pre,
 reprint,
 amsmath,amssymb,
 superscriptaddress,
 twocolumn
]{revtex4-2}
\usepackage{graphicx}
\usepackage{MnSymbol}
\usepackage{xcolor}
\usepackage{outlines}
\usepackage{CJK}

\usepackage{dcolumn}
\usepackage{bm}
\usepackage{mathtools}
\usepackage{amssymb,amsmath,amsfonts,dsfont}
\usepackage{hyperref}
\usepackage{color}
\usepackage{float}

\begin{document}

\begin{CJK*}{UTF8}{mj}
\title{Statistical mechanics in continuous space with tensor network methods}
\author{Gunhee Park (박건희)}
\thanks{These two authors contributed equally}
\affiliation{Division of Engineering and Applied Science, California Institute of Technology, Pasadena, California 91125, USA}

\author{Tomislav Begu\v{s}i\'{c}}
\thanks{These two authors contributed equally}
\affiliation{Division of Chemistry and Chemical Engineering, California Institute of Technology, Pasadena, California 91125, USA}

\author{Si-Jing Du}
\affiliation{Division of Engineering and Applied Science, California Institute of Technology, Pasadena, California 91125, USA}

\author{Johnnie Gray}
\affiliation{Division of Chemistry and Chemical Engineering, California Institute of Technology, Pasadena, California 91125, USA}

\author{Garnet Kin-Lic Chan}
\email{gkc1000@gmail.com}
\affiliation{Division of Chemistry and Chemical Engineering, California Institute of Technology, Pasadena, California 91125, USA}
\date{\today}

\begin{abstract}
Tensor network (TN) methods are well established for computing partition functions in statistical mechanics, though this use has traditionally been limited to lattice models. We extend the scope of TN methodology to interacting particle systems in continuous space. Through a real-space discretization combined with a cell-based coarse-graining scheme, we formulate an effective lattice model that explicitly preserves spatial locality. The partition function of this model is represented as a TN, and the thermodynamic quantities are computed via boundary contraction. We apply this framework to the two-dimensional hard-disk problem and demonstrate the strengths of the TN formulation compared to existing Monte Carlo simulations.
\end{abstract}

\maketitle
\end{CJK*}

\section{Introduction}

Tensor networks (TN) provide a powerful framework for the efficient representation and manipulation of high-dimensional objects in both quantum many-body physics and classical statistical mechanics~\cite{Verstraete01032008, Orus2019, CiracRMP2021, OkunishiNishino2022, ran2020tensor, Xiang_2023}. In classical statistical mechanics, the primary applications of this framework have been in the context of lattice models, such as the classical Ising or XY models in two or three dimensions~\cite{Nishino1995, Nishino1997, Nishino2001, LevinNave2007, Orus2008tebd, Xie2009srg, Xie2012hotrg, YuXie2014XYmodel, EvenblyVidal2015tnr, YangGuWen2017}. Unlike traditional Monte Carlo (MC) methods, the TN approach provides a deterministic route to calculating thermodynamic quantities. They 
thus avoid the problem of large-scale sampling, which is replaced by the computational task of tensor contraction, with different complexity characteristics. In addition, TN approaches allow for new kinds of simulations. For example, using infinite TN, one can estimate results in the thermodynamic limit without using size extrapolation, while TN computation also allows us to directly obtain the absolute free energy, a major challenge in MC techniques.

This work aims to extend the scope of the TN methods in classical statistical mechanics to interacting particle systems in a continuous space. While existing TN literature on statistical mechanics in continuous spaces has employed a ``particle" picture~\cite{DEKTOR2021110295, HUR2023101575, CHEN2023111646, TANG2024113110, Grimm2026integraldecimation, johnson2025tensornetworksliquidsheterogeneous}, where the low-rank structure of TN captures inter-particle correlations, this work instead seeks to exploit spatial locality, including high-dimensional cases, analogous to the approaches conventionally used in lattice models.
To achieve this, we formulate an effective lattice model through a real-space discretization combined with a cell-based coarse-graining scheme. A TN representation is then constructed for the partition function of this effective lattice model. 
Thermodynamic quantities are subsequently computed through the contraction of the resulting tensor network.

We apply this methodology to the statistical mechanics of a two-dimensional (2D) interacting particle system, namely particles with a hard-disk (HD) potential. 
Local observables, such as densities and pair correlation functions, are computed and benchmarked against results from standard Markov chain Monte Carlo (MCMC) simulations, using the same real-space discretization. We use the infinite TN framework to describe the thermodynamic limit, as well as finite TN simulations to better capture the liquid-to-solid transition.
Finally, we report the absolute free energies from the TN calculations, where we observe the complexity of the computation scales linearly with system size, in stark contrast to the approximately exponential scaling required by the Wang-Landau algorithm~\cite{wang2001prl, wang2001determining}. We conclude with an outlook on the future development of these techniques.

\section{Methods}

\subsection{Partition function expression}

\begin{figure}[t]
    \centering
    \includegraphics[width=\columnwidth]{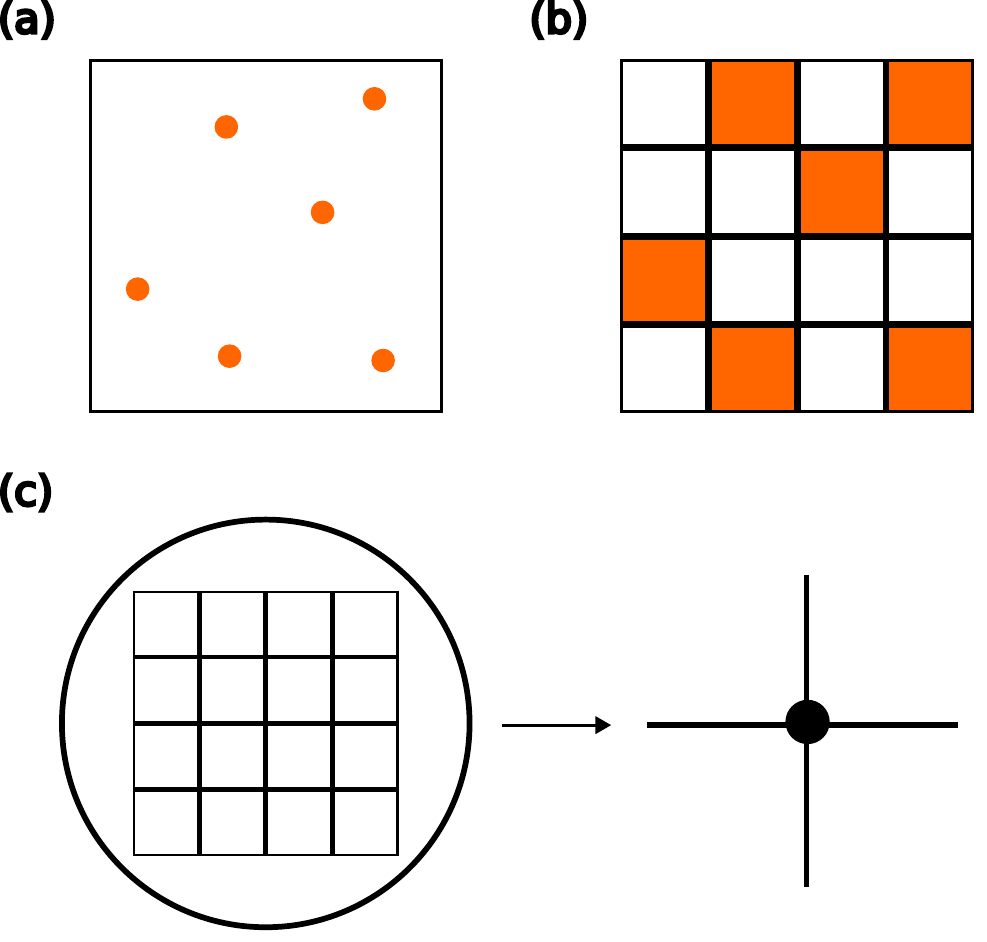}
    \caption{(a) Particles in a box. (b) Same system in the site representation, where colors denote occupied lattice grid points. (c) Grid points in a cell are coarse-grained into one effective site. }
    \label{fig:schematic1}
\end{figure}

Our objective is to compute the grand canonical partition function $\Xi$, which is defined as:
\begin{equation}
    \Xi(\mu) = \sum_{N} e^{\beta \mu N} Z_N.
\end{equation}
Here, $Z_N$ is the canonical partition function for a system containing $N$ particles:
\begin{equation}
    Z_N = \frac{1}{N!} \int d\bm{r}_1 \cdots d\bm{r}_N e^{-\beta U(\bm{r}_1, \cdots, \bm{r}_N)}.
\end{equation}
$\beta$ and $\mu$ are the inverse temperature and the chemical potential, respectively, and the total potential energy $U$ is the sum of all pairwise potentials $V(|\bm{r}_i - \bm{r}_j|)$ between every pair of particles $i$ and $j$,
\begin{equation}
    U(\bm{r}_1, \cdots, \bm{r}_N) = \sum_{1 \leq i < j \leq N} V(|\bm{r}_i - \bm{r}_j|).
\end{equation}
To compute the partition function using TN methods, we first discretize the continuous spatial coordinates $\bm{r}$ into a finite set of discrete lattice grid points, which we denote as $G$. Upon discretization, the canonical partition function $Z_N$ transforms into a sum:
\begin{equation}
    Z_N = \frac{1}{N!} (\Delta v)^N \sum_{\bm{r}_1 \in G}  \cdots \sum_{\bm{r}_N \in G} e^{-\beta U(\bm{r}_1, \cdots, \bm{r}_N)},
\end{equation}
where $\Delta v$ refers to the volume occupied by each grid point. For instance, in a 2D uniform grid with a lattice constant $a$, $\Delta v = a^2$. 
This particular expression, which involves summing over the coordinates of the $N$ particles spanning the entire volume, is referred to as the ``particle" representation.
In contrast, TN methods often perform more efficiently when defined using local site coordinates, similar to conventional lattice statistical mechanics models (e.g., the classical Ising model). We refer to this alternative as the ``site" representation.
Within this representation, the grand canonical partition function $\Xi$ is now expressed as,
\begin{equation}
    \Xi(\mu) = \sum_{\{n_{k}\} \subset \{0, 1\}^{|G|}} \left(e^{\beta \mu} \Delta v \right)^{\sum_k n_{k}}  e^{-\beta U(\{n_{k}\})},
\end{equation} 
where each $k$ refers to a lattice grid site in $G$, and $n_{k}$ refers to an occupation number at site $k$. The total potential energy is given by,
\begin{equation}
    U(\{n_{k}\}) = \sum_{1 \leq k < k' \leq |G|} V(|\bm{r}_k - \bm{r}_{k'}|) n_{k} n_{{k'}} .
\end{equation}

\subsection{Cell-based coarse-graining scheme} \label{sec:coarse-graining}

Instead of using the fine grid as a basis for TN, we introduce a coarse-graining scheme related to the cell model, a popular approach between the 1930s and 1950s~\cite{HirschfelderEyring1937, LennardJonesDevonshire1937, Kirkwood1950, barker1963lattice}. This approach applies to systems described by a pairwise potential with short-range repulsion, a generic feature found in nearly all interatomic or intermolecular potentials.

Intuitively, two sites within the range of a strong repulsive potential cannot be occupied simultaneously.
Specifically, if $V(r)>V_c$ for distances $r<r_c$, where $V_c$ is a sufficiently large energy threshold, configurations where two sites $k$ and $k'$ are both occupied ($n_k = n_{k'} = 1$) at a separation $|\mathbf{r}_k - \mathbf{r}_{k'}| < r_c$ become energetically suppressed, i.e., their probability of occurrence is effectively zero.

Based on the above observation, we partition the sites into a cell such that the maximum distance between any two sites within the cell is smaller than $r_c$ (Fig.~\ref{fig:schematic1}c). Within each cell, we restrict the allowed configurations to those that are either entirely unoccupied (all-zero) or have exactly one occupied site. This constraint significantly reduces the configuration space. For example, a 4-site cell allows $1+4 = 5$ configurations:
\[
(0, 0, 0, 0), (1, 0, 0, 0), (0, 1, 0, 0), (0, 0, 1, 0), (0, 0, 0, 1).
\]
This is a reduction from the full $2^4=16$ possible configurations. In this work, we consider only these singly occupied configurations, but it is straightforward to extend this approach to multiply occupied configurations.

We treat each coarse-grained cell as a single, effective local ``site". This coarse-graining procedure effectively reduces the connectivity (or effective range) of the pairwise potentials. This simplification will be beneficial for the TN methods to be introduced in the next section.

\subsection{Tensor networks for partition function}

\begin{figure}[t]
    \centering
    \includegraphics[width=\columnwidth]{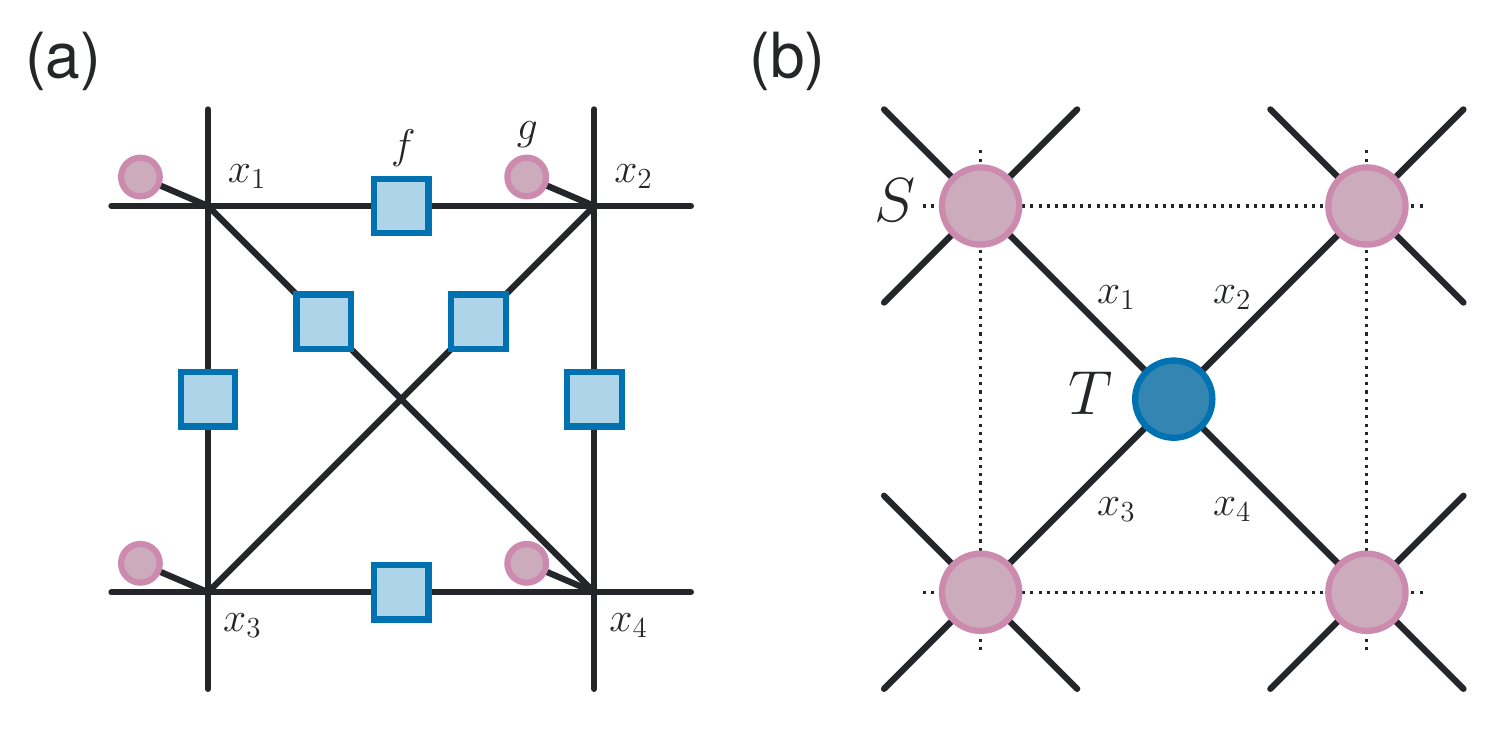}
    \caption{(a) Diagrammatic representation of the factor graph for the hard disk partition function, where the on-site and pairwise terms are shown as pink circles and blue squares, respectively. (b) When the pair connectivity is restricted to nearest-neighbor and next-nearest-neighbor interactions, the factor graph can be transformed into an equivalent representation where tensors are placed at the original lattice sites and at the centers of the square units. }
    \label{fig:fig-tn}
\end{figure}

In the language of TN methods, the grand canonical partition function $\Xi$ can be expressed in terms of a factor graph~\cite{Mezard_Montanari_book_2009},
\begin{equation}
    \Xi = \sum_{\{ x_k \} } \prod_{k < k'} f_{kk'}(x_k, x_{k'}) \prod_k g_k(x_k),
\end{equation}
where $x_k$ is a local configuration within the coarse-grained site $k$. $g_k$ and $f_{kk'}$ are given by:
\begin{equation}
g_k (x_k) = 
\begin{cases} 
e^{\beta \mu} \Delta v & \text{one occupied } x_k \\
1 & \text{unoccupied } x_k  
\end{cases}
\end{equation}
\begin{equation}
f_{kk'} (x_k, x_{k'}) = 
\begin{cases} 
e^{-\beta V(x_k, x_{k'})} & x_k, x_{k'} \text{ are both occupied}  \\
1 & \text{otherwise} 
\end{cases}
\end{equation}
where $V(x_k, x_{k'})$ corresponds to the pair potential between the corresponding occupied sites at each cell. Its diagrammatic representation is displayed in Fig.~\ref{fig:fig-tn}a, where the on-site terms $g_k$ and pairwise terms $f_{kk'}$ are represented by pink circles and blue squares, respectively.

For an arbitrary pairwise potential, the corresponding factor graph has all-to-all connectivity. 
In this work, we restrict our focus to short-range interactions where the potential only acts between the nearest-neighbor (NN) and next-nearest-neighbor (NNN) cells. We emphasize that this does not restrict interactions to NNN on the original fine grid.
Rather, it signifies that the interaction range on the fine grid remains within the linear dimensions of the coarse-grained cells.

Rather than turning the factor-graph into a TN with the same connectivity, we follow the scheme in Ref.~\cite{LiYang2021j1j2} for partition functions with NNN interactions, which leads to a 2D TN defined on a square lattice.
Therein, tensors are placed both at the original lattice sites and at the centers of the square units, denoted by tensors $S$ and $T$ in Fig.~\ref{fig:fig-tn}b, respectively.
The tensor $S$ is given by contracting the on-site tensor $g$ to the identity tensor, i.e., $S(x_1, x_2, x_3, x_4) = g(x_1) \delta_{x_1,x_2, x_3, x_4}$, and the tensor $T$ is given by,
\begin{equation}
    T = f_{14} \cdot f_{23} \cdot \sqrt{f_{12}} \cdot  \sqrt{f_{13}} \cdot  \sqrt{f_{24}} \cdot  \sqrt{f_{34}},
\end{equation}
where the square root is applied elementwise, and the coordinates 1, 2, 3, 4 are arranged as in Fig.~\ref{fig:fig-tn}.
This construction results in a new square lattice that is rotated by $\pi/4$ relative to the original lattice. In the resulting TN, the bond dimension $D$ corresponds to the number of possible configurations per cell.

\subsection{Tensor network contraction}

\begin{figure}[t]
    \centering
    \includegraphics[width=\columnwidth]{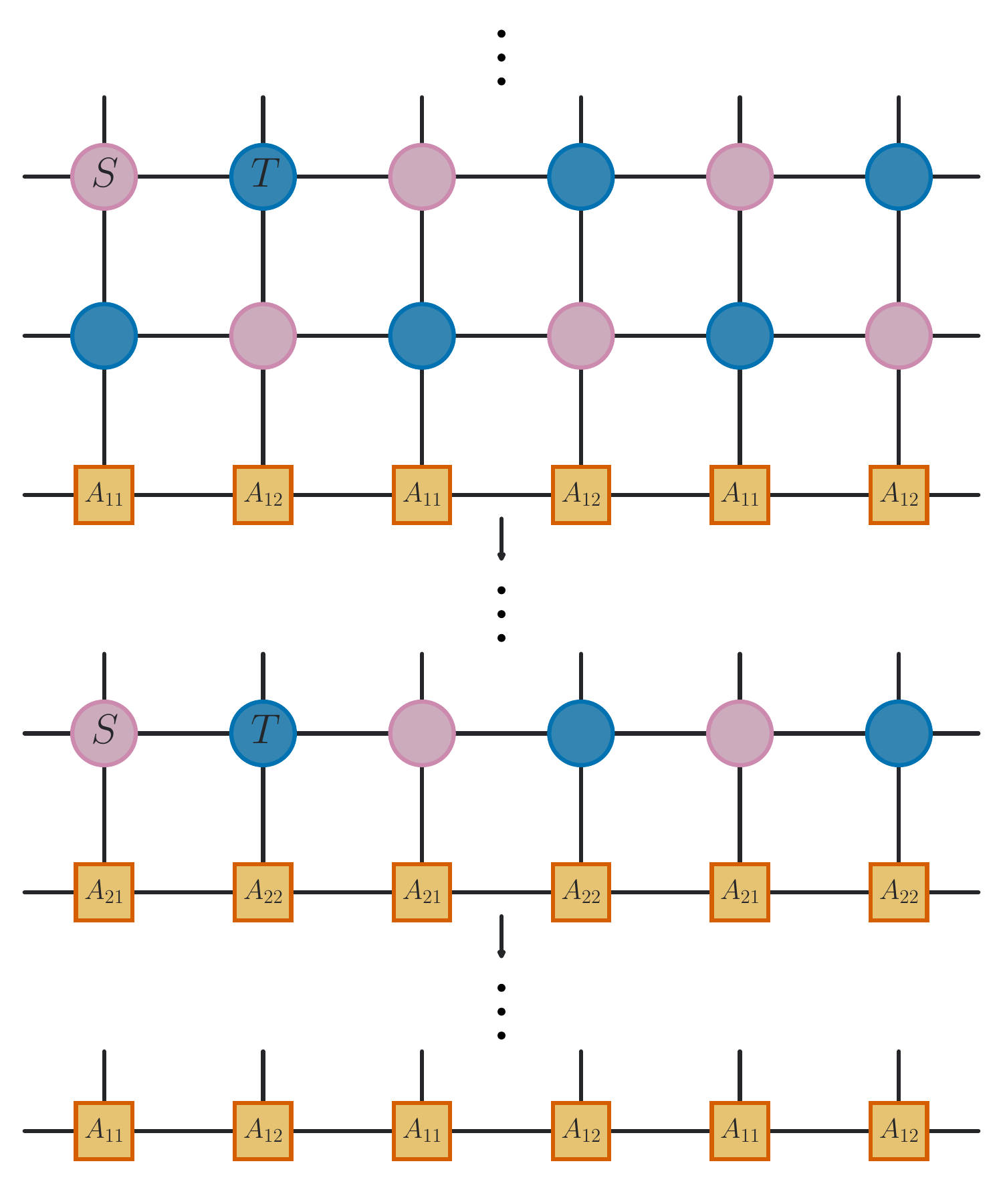}
    \caption{$2\times2$ unit cell structure for the transfer matrix of the partition function composed of $S$ and $T$ tensors in the infinite tensor network. Its boundary MPS is characterized by two alternating two-site uniform MPS with tensors $(A_{11}, A_{12})$ and $(A_{21}, A_{22})$. }
    \label{fig:fig-infinite-tn}
\end{figure}

We adopt the boundary contraction strategy using MPS ~\cite{Jordaorus2008, Verstraete01032008, Orus2008tebd} to contract the 2D TN. We apply the boundary contraction method under two settings: (1) an infinite-system limit and (2) a finite box with open boundary conditions (OBC). The infinite-system limit is addressed by the boundary contraction method within the infinite TN framework, which directly computes thermodynamic limit results. The boundary contraction is taken along the $\pi/4$-rotated angle from the TN construction of the NNN interaction. The boundary MPS in the infinite TN is represented as a uniform MPS~\cite{Haegeman2013, Haegeman2017annurev, ZaunerStauber2018, Fishman2018, Vanderstraeten2019SciPostLectNotes}.
The checkerboard arrangement of the $S$ and $T$ tensors defines a $2 \times 2$ unit cell for the transfer matrix of the partition function. Consequently, the boundary MPS is characterized by two alternating uniform MPS, each possessing a two-site unit cell with tensors $(A_{11}, A_{12})$ and $(A_{21}, A_{22})$~\cite{Nietner2020efficient} in Fig.~\ref{fig:fig-infinite-tn}.
The power iteration of the multiplication of the uniform matrix product operator (MPO) and MPS~\cite{Vanhecke2021} is carried out until the boundary MPS converges.  
For the finite OBC simulations, a successive randomized compression algorithm~\cite{Camano2026successive} is used for the MPO-MPS contraction. For both finite and infinite algorithms, the computational complexity is $\mathcal{O}(D^2 \chi^3 + D^4 \chi^2)$ \cite{Vanhecke2021, Camano2026successive} where $\chi$ is the bond dimension of the boundary MPS.

\section{Results}

In this work, a hard-disk (HD) pairwise potential is considered:
\begin{equation}
V_{\text{HD}}(r) = 
\begin{cases} 
0 & \text{if } r \geq \sigma \\
+\infty & \text{if } r < \sigma 
\end{cases}
\end{equation}
We work in dimensionless units where $\sigma = 1$ and $\beta=1$. Since the statistical ensemble for the HD potential is independent of temperature, the value of $\beta$ does not change the result, i.e., the system is athermal.
We use grand canonical Monte Carlo (GCMC) based on Markov chain sampling with the Metropolis-Hastings algorithm as a reference method to compute densities and pair correlation functions.

For the construction of the cells, the cell width is set to be $l_c = \frac{7}{8} \sigma$, and each cell covers a $6 \times 6$ fine grid, resulting in a fine grid lattice discretization of $a = \frac{1}{6} l_c  = \frac{7}{48} \sigma \approx 0.148534 \sigma$. The corresponding TN bond dimension is $D=37$.
Convergence to the continuum limit is demonstrated via the GCMC results provided in the Appendix~\ref{sec:lattice-convergence}, which compares the GCMC density in a continuous space simulation versus the lattice. 
With this setting, the TN from the HD potential can be constructed using NN and NNN interaction terms.

\begin{figure}[t]
    \centering
    \includegraphics[width=0.9\columnwidth]{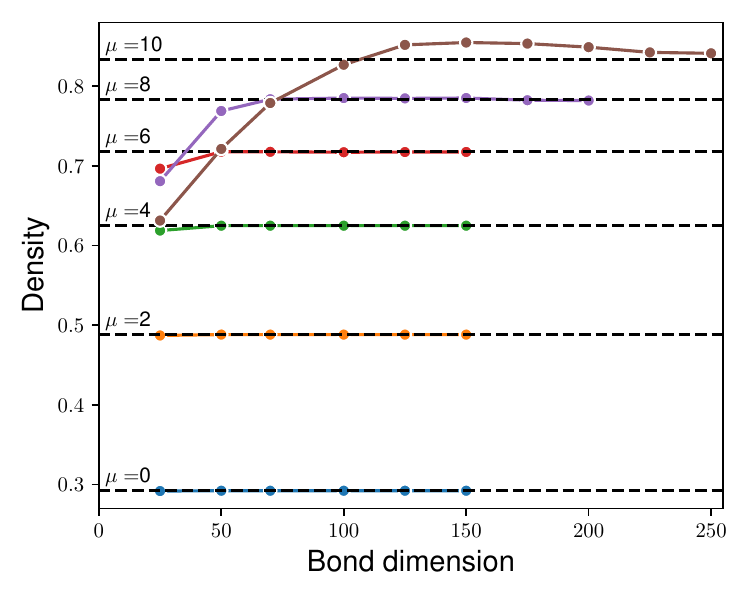}
    \caption{Within the infinite TN framework, the density is plotted as a function of MPS bond dimension using the boundary contraction method. The reference MC data is shown as the dashed lines, where the MC is performed on a finite system with $L=28$ using periodic boundary conditions.}
    \label{fig:infinite-tn2d-density}
\end{figure}

First, we demonstrate the infinite TN results. In Fig.~\ref{fig:infinite-tn2d-density}, the density is plotted as a function of the boundary MPS bond dimension $\chi$ at various chemical potentials. The reference lattice GCMC results are added as dashed lines, which are performed in a finite system with $L=28$ and periodic boundary conditions (PBC). The finite-size effect at $L=28$ is smaller than $10^{-4}$ by comparing to results from $L=14$. At low and intermediate density regimes ($\mu = 0,2,$ and $4$), the TN results converge rapidly at low bond dimensions. As the chemical potential and density increase, larger bond dimensions are required for convergence. In most cases, the values converge to the MC reference values. However, the $\mu=10$ results show slow convergence to the reference MC result.

We observe that the boundary MPS of the infinite TN does not converge rapidly with the power method when $\mu$ is larger than 10.
We assume this stems from the proximity to the liquid-to-solid phase transition.
In this regime, the physical system spontaneously breaks translational invariance, yet it is represented here by a translationally invariant boundary MPS with a finite bond dimension. Thus, the MPS describes a translational superposition of solid phases, requiring a larger bond dimension to achieve convergence.
In general, the infinite TN result can be improved by choosing the unit cell in the boundary MPS to be commensurate with the emergent order~\cite{corboz2011stripes}.

\begin{figure}[t]
    \centering
    \includegraphics[width=\columnwidth]{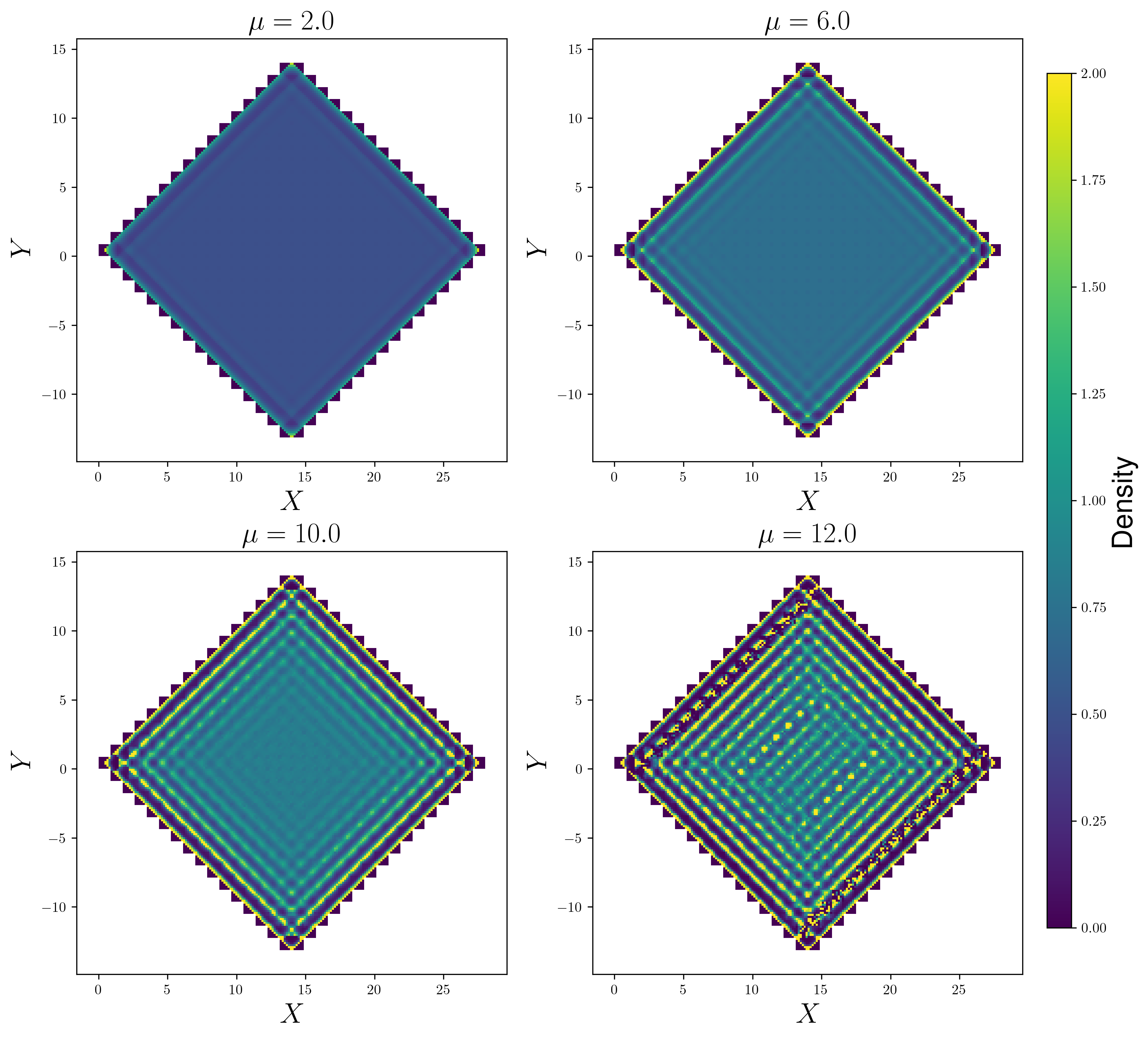}
    \caption{Spatially resolved densities in a finite box with open boundary conditions at $\mu=2,6,10,$ and 12.}
    \label{fig:density-finite}
\end{figure}

\begin{figure*}[t]
    \centering
    \includegraphics[width=\textwidth]{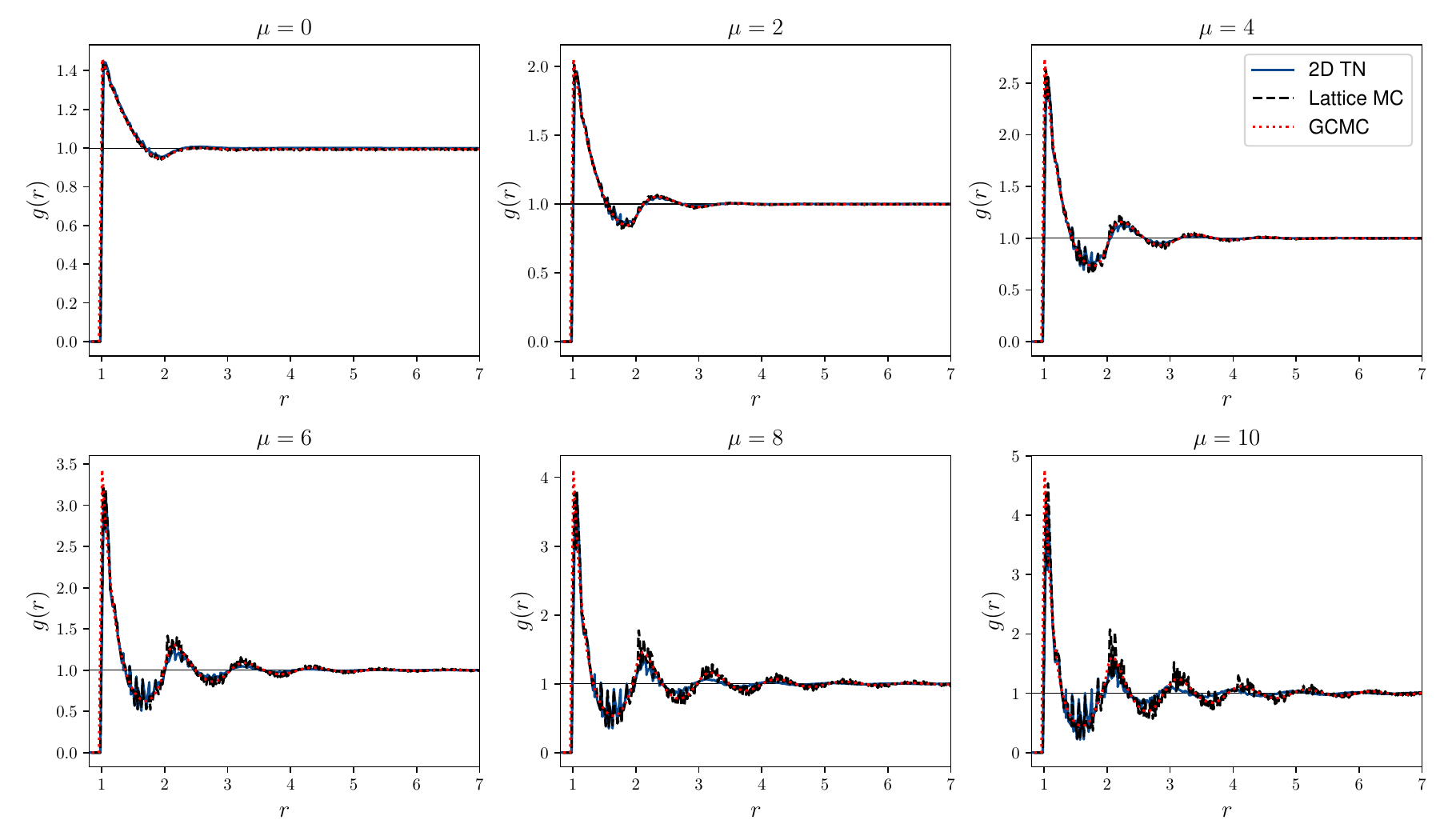}
    \caption{Pair correlation functions computed from infinite TN and MC calculations. For the MC reference, the GCMC data from both the continuous space (red dotted) and the discretized lattice (black dashed) are presented.}
    \label{fig:tn2d-gr}
\end{figure*}

Another way to describe the broken symmetry phase is to use a finite TN where the boundary MPS is free from the translational invariance constraint.
For technical reasons, OBC TN simulations are easier than PBC simulations, and thus, we demonstrate finite OBC calculations below to capture the liquid-to-solid phase transition. 
The two-dimensional liquid-to-solid phase transition has been an active area of research from the early days of computer simulations~\cite{AlderWainwright1957, Strandburg1988rmp} until today~\cite{Bernard2011, Engel2013, LiKrauth2022harddisk}.
In Fig.~\ref{fig:density-finite}, spatially resolved densities from such finite box simulations are plotted for $\mu=2,6,10,$ and 12 within a diamond-shaped box containing 16 coarse-grained cells along each edge, corresponding to a $L \approx 19.2$ finite box.
To eliminate the artificial staircase structure of the diagonal boundaries, an infinite-potential wall is introduced to create a smooth confining edge.
In these OBC calculations, we use a bond dimension of up to $\chi=500$ for $\mu=12$. At $\mu=12$, a crystal forms, although not in the bulk hexagonal crystal structure due to the finite open boundary, which serves as a source of frustration. Nonetheless, these results illustrate that the TN simulations can capture a liquid-to-solid phase transition.

In Fig.~\ref{fig:tn2d-gr}, we compute pair correlation functions, $g(\bm{r}_1,\bm{r}_2) = P_2(\bm{r}_1,\bm{r}_2) / (P_1(\bm{r}_1) \cdot P_1(\bm{r}_2) )$, where $P_1$ and $P_2$ are 1-site and 2-site marginal probabilities, respectively. In the continuum limit, $g(\bm{r}_1,\bm{r}_2)$ becomes a function of distance $r=|\bm{r}_1 - \bm{r}_2|$ with spherical symmetry. In our cell model, the spherical symmetry is not strictly satisfied. We fix the direction of $\bm{r}_1 - \bm{r}_2$ along a row of the 2D TN, and collect the data on $g(r)$ along this direction. (Data in Fig.~\ref{fig:tn2d-gr} is obtained from the infinite TN simulation). As a reference, the GCMC results from both the continuous space (red dotted) and the discretized lattice (black dashed) are plotted. Compared with the GCMC results in continuous space, the TN and lattice GCMC results exhibit additional high-frequency oscillations but still capture the overall envelope of the pair correlation function, illustrating the increasing long-range order.

\begin{figure}[t]
    \centering
    \includegraphics[width=\columnwidth]{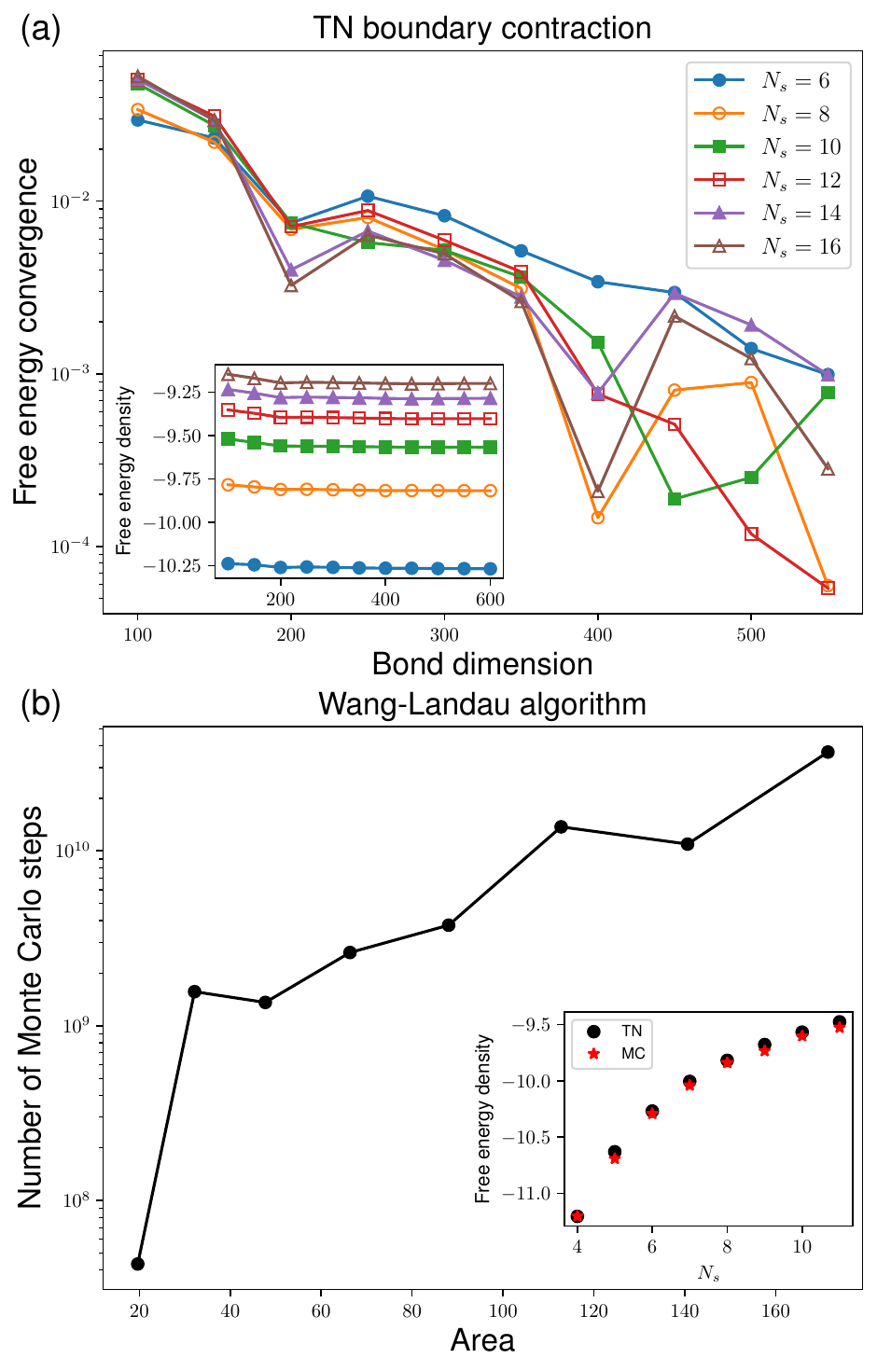}
    \caption{(a) Free energy convergence using the TN boundary contraction method as a function of MPS bond dimension for various system sizes. Free energy convergence is quantified by the computed free energy density difference between the given bond dimension and the maximum bond dimension, $\chi_{\text{max}}=600$. Free energy density values for different system sizes are plotted in the inset. (b) The number of Monte Carlo steps in the Wang-Landau algorithm as a function of area. The inset shows a comparison of the free energy density calculated via the TN and MC methods. See the main text for the details of the Monte Carlo procedures. }
    \label{fig:free_energy_convergence}
\end{figure}

Beyond calculating observables, TN methods provide a natural framework for computing absolute free energy. Since the TN directly represents the partition function, its TN contraction cost scales linearly with the system size for a fixed boundary MPS bond dimension.  Fig.~\ref{fig:free_energy_convergence}a shows the convergence of the free energy density, $F = - \frac{1}{A} \log \Xi$, where $A$ is the total area of the system, for various bond dimensions and system sizes with $N_s $  cells along each edge at $\mu=12$. The maximum bond dimension used here is $\chi_{\text{max}} = 600$, and the convergence is quantified by the free energy density difference from the bond dimension $\chi_{\text{max}}$, i.e., $\Delta F = |F_{\chi} - F_{\chi_{\text{max}}}|$. The results show that the bond dimension required to achieve a target accuracy remains independent of the system size, verifying the linear scaling cost of TN contraction.

In contrast, MC methods face significant challenges in obtaining absolute free energies. The common practice is to compute the relative free energy from reference points or to compute the absolute free energy using the flat histogram technique~\cite{frenkel2023understanding}. The former includes thermodynamic integration and free energy perturbation~\cite{frenkel2023understanding, kirkwood1935fluid, zwanzig1954perturbation, christ2010review}, and the latter includes the Wang-Landau (WL) algorithm~\cite{wang2001determining, wang2001prl, Yan2002dos, Dayal2004, Singh2012review}. Since the former depends on the choice of reference point, we choose the WL algorithm as a reference for comparison for the cost of free energy calculation in the hard disk problem.

The WL algorithm estimates the density of states as a function of the number of particles, $\rho (N)$, by performing MCMC where the transition probabilities are inversely proportional to the current $ \rho (N)$ estimate. The algorithm iteratively updates $ \rho (N)$ by a modification factor $f$ until a flat histogram of visited states is achieved. This process is repeated while reducing $f$ via $f_{i+1} = \sqrt{f_i}$ until the density of states converges. The absolute free energy can be calculated from the final density of states.

The computational cost of the above algorithm for the hard disk problem is quantified as follows. Starting with $f_0 = e$, $f$ is updated for 26 iterations. We define the computational cost as the number of Monte Carlo steps required in the final iteration to satisfy the flat histogram criterion. For statistical reliability, the algorithm is first evolved through 25 iterations, and subsequently, 100 independent realizations are performed for the final iteration alone. The cost is then calculated as the geometric mean of the number of Monte Carlo steps across these independent runs. Here, we target problems with $\mu=12$, so we truncate high-$N$ states that are not contributing to the free energy based on the density of states estimate after 25 iterations when computing the number of Monte Carlo steps at the final iteration. Consistent with the TN setup, we perform the Markov chain using the discretized coarse-grained cell representation and using single-site updates.

Fig.~\ref{fig:free_energy_convergence}b illustrates the computational cost of the WL algorithm as a function of the system size expressed as the area. The cost increases rapidly with area, exhibiting approximately exponential growth.
The inset shows a direct comparison of the free energy density calculated via the TN and MC methods. The MC method provides free energy density values consistent with the TN results, although they do not match the high numerical accuracy of the converged TN method.
This trend underscores the efficiency of the TN method, where it is possible to reach high accuracy while maintaining linear scaling with system size.

\section{Conclusions}

In this work, we described the TN simulation of statistical mechanics problems in a continuous space. By using real-space discretization and a cell-based coarse-graining scheme, we developed a way to map interacting particle systems onto effective lattice models while keeping spatial locality. Our results for the two-dimensional hard-disk potential confirm that this deterministic TN contraction approach accurately reproduces local observables, such as densities and pair correlation functions, when benchmarked against the standard Monte Carlo simulations. Notably, we demonstrated that some of the unique features of TN simulations, such as the possibility of infinite system simulations and computation of the absolute free energy, can be realized in our representation of continuous statistical mechanical problems.

While this work focuses on two-dimensional systems, the method can be extended to three dimensions using 3D TN, as has already been done for various lattice models~\cite{Nishino2001, Xie2012hotrg}. We are currently working to expand the framework from short-range to longer-range potentials and will report these findings in future work. With these developments, we anticipate that this approach will become a versatile tool for studying complex continuous systems in equilibrium, non-equilibrium, and quantum statistical mechanics.

\begin{acknowledgments}
The authors thank David Limmer for helpful discussions.
This work was supported by
the US Department of Energy, Office of Science, Accelerated Research in Quantum Computing Centers, Quantum Utility through Advanced Computational Quantum Algorithms, through Award No. DE-SC0025572. GKC is a Simons Investigator in Physics.
\end{acknowledgments}

\appendix

\section{Lattice discretization convergence}
\label{sec:lattice-convergence}

\begin{figure}[H]
    \centering
    \includegraphics[width=\columnwidth]{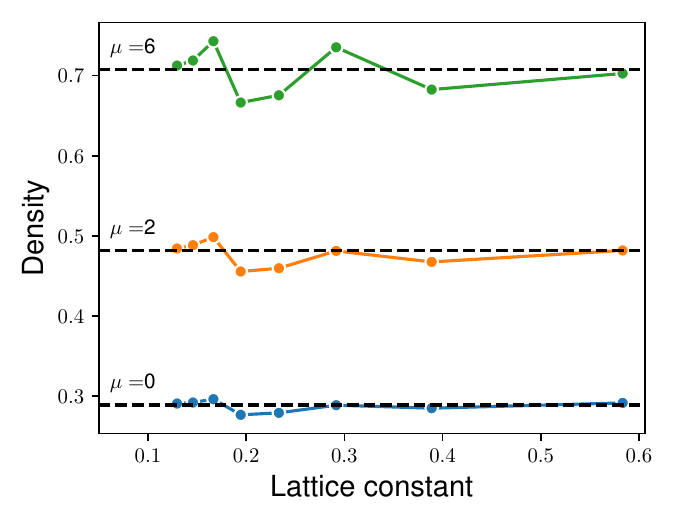}
    \caption{Density as a function of lattice discretization constant compared to density in the continuum, indicated as dashed lines, at three different chemical potential values.}
    \label{fig:lattice-convergence}
\end{figure}

Fig.~\ref{fig:lattice-convergence} shows the convergence of the density values as a function of the lattice discretization in the lattice GCMC calculations compared to GCMC in the continuum, where the smallest lattice discretization used in the plot corresponds to the lattice discretization value used in this work.

\bibliography{references}

\end{document}